# Direct Estimation of Firing Rates from Calcium Imaging Data

Elad Ganmor[1], Michael Krumin[2], Luigi F. Rossi[2], Matteo Carandini[2] & Eero P. Simoncelli[1,3]

1. Center for Neural Science, New York University
2. UCL Institute of Ophthalmology, University College London
3. Howard Hughes Medical Institute, New York University

Corresponding author:

    Elad Ganmor
    New York University
    Center for Neural Science
    4 Washington Place, Room 1027
    New York, NY 10003
    (212) 992-8752
    elad.ganmor@gmail.com

## *Abstract*

Two-photon imaging of calcium indicators allows simultaneous recording of responses of hundreds of neurons over hours and even days, but provides a relatively indirect measure of their spiking activity. Existing "deconvolution" algorithms attempt to recover spikes from observed imaging data, which are then commonly subjected to the same analyses that are applied to electrophysiologically recorded spikes (e.g., estimation of average firing rates, or tuning curves). Here we show, however, that in the presence of noise this approach is often heavily biased. We propose an alternative analysis that aims to estimate the underlying rate directly, by integrating over the unobserved spikes instead of committing to a single estimate of the spike train. This approach can be used to estimate average firing rates or tuning curves directly from the imaging data, and is sufficiently flexible to incorporate prior knowledge about tuning structure. We show that directly estimated rates are more accurate than those obtained from averaging of spikes estimated through deconvolution, both on simulated data and on imaging data acquired in mouse visual cortex.

## *Introduction*

Neurons convey information using spikes. For example, sensory neurons emit different numbers of spikes with different timings in response to different stimuli. Yet these responses are often described as 'noisy' since the responses to identical stimuli vary from one trial to the next even when external factors are carefully controlled. Therefore, it is common to summarize neural responses in terms of average spike counts across multiple trials, acknowledging the stochastic nature of spike generation.

Estimating firing rates is a ubiquitous form of analysis throughout neurophysiology. Well-known examples include spike count histograms and tuning curves. The spike count histogram represents the average response over a fixed time interval across many repeats of the same experimental condition, relative to some stimulus or behavioral response, while the tuning curve is a measure of the firing rate under different experimental conditions (typically, stimuli or actions that vary along some parametric axis). Estimating rates from spikes thus requires combining spike counts across repeated measurements, and is relatively straightforward when spikes are recorded directly, as is the case in electrophysiology experiments.

Imaging techniques provide an appealing means of measuring neural activity across populations. Two-photon imaging of calcium indicators, in particular, allows one to measure up to thousands of neurons simultaneously at single-cell or even sub-cellular resolution. Moreover, imaging techniques are readily combined with genetic and opto-genetic methods to record and stimulate specific cell types. But in comparison to electrical recordings, calcium imaging provides a less direct measure of spiking activity: the acquired image represents the intensity of a fluorescence signal that depends on the intracellular calcium level, which, in turn, is driven by the spiking activity. Consequently, it is not straightforward to estimate firing rates from calcium imaging data -- simple averaging across trials is not sufficient.

The most intuitive method of obtaining spike rates from imaging data is to invert the two steps of the observation process: estimating the spike trains from the fluorescence, and then averaging the estimated spike counts to infer the rate. Estimating spike trains from fluorescence is commonly referred to as deconvolution, reflecting an assumption that intracellular calcium levels are the outcome of a convolution of the spike train with a known temporal filter - deconvolution seeks to undo this process. Several algorithms have been proposed to perform this deconvolution step (Vogelstein et al. 2010; Oñativia, Schultz, and Dragotti 2013; Dyer et al. 2010; Grewe et al. 2010), and have been successfully applied to estimate spike rates and tuning curves from imaging data (Smith and Häusser 2010; Ko et al. 2013).

Despite the successes of deconvolution methods, it is important to recognize that spike counts estimated from the deconvolution process are only approximate. Moreover, as we show, errors can be substantial, and more importantly may depend systematically on the firing rate. As such, these errors are not reduced as one would expect by averaging over repeated trials. Furthermore, deconvolution methods do not take into account the structure of the experiment (specifically, the timing and sequence of stimuli or behavioral responses), although these factors may have a significant impact on spiking activity. And even if subsequent stages of analysis attempt to incorporate such experimental details, the decisions made during the deconvolution stage typically cannot be undone: missed spikes are irretrievably lost and falsely detected spikes cannot be distinguished from their correctly identified neighbors.

Here, we introduce an alternative approach to estimating rates, spike count histograms, or tuning curves directly from calcium fluorescence measurements, without the need for deconvolution. This direct approach mitigates the biases associated with sequential estimation schemes (deconvolution followed by averaging or tuning curve estimation), especially when measurements are noisy. Our method operates by maximizing the likelihood of a simple model for the fluorescence generation, integrating over the distribution of unobserved spike counts. We demonstrate the effectiveness of this direct estimation method on both model-simulated data sets, and real calcium imaging data, including data for which ground-truth spiking activity was obtained with simultaneous electrophysiological measurements.

## Methods

Our estimation procedure is derived from an observation model that expresses the relationship between the calcium fluorescence signal and the underlying firing rate. Each component of this observation model is simple and has appeared in the literature, but the particular combination and methodology we present is, to our knowledge, novel.

## Generative model for imaging data

A graphical diagram of the model is shown in Fig. 1. We assume spikes arise from an inhomogeneous Poisson process, with an unknown rate. The rate is either assumed to be constant for the duration of each experimental condition, or expressed as a parametric function of external covariates (e.g., sensory stimuli, or behavioral responses). We also assume that each spike causes an instantaneous rise in calcium level followed by an exponential decay to baseline (Vogelstein et al. 2010; Oñativia, Schultz, and Dragotti 2013; Smith and Häusser 2010). Moreover, we assume that the calcium arising from each incoming spike is additive. Thus, the calcium signal arises from the spike train via convolution with an exponentially decaying filter. Finally, we assume that the fluorescence measured by the microscope is a scaled version of the calcium level, corrupted with additive Gaussian noise. The latter assumption is appropriate because calcium signals from neurons are typically averaged over multiple pixels. Even if the noise of each pixel were better described as Poisson due to the nature of photons, the noise in their sum, per the Central Limit Theorem, is approximately Gaussian (Wilt, Fitzgerald, and Schnitzer 2013). For smaller structures such as spines, however, this assumption may need to be replaced with one of Poisson noise.

## Fitting the model

We fit the model by finding the firing rate $\lambda(t)$ that maximizes the likelihood of the observed fluorescence trace $F(t)$. More generally, given a stimulus or behavioral response $S(t)$, we find the parameters $\theta$ governing its mapping into firing rate, $\lambda(S(t), \theta)$. We treat our data as discretely sampled in time intervals of $\Delta t$, the sampling rate of the experimental measurements. We denote the vector of all samples up to time $T$ by $F(0 \dots T)$, and likewise for the rate $\lambda(0 \dots T)$. Since the probability of the observed fluorescence depends only on the instantaneous firing rate, and on the value at the previous time step (due to the exponential decay of the calcium level), and since the transformations from stimuli to firing rate, and from spikes to calcium are assumed to be deterministic, the likelihood function may be factorized:

(1)
$$P(F(0 \ldots T)|\lambda(0 \ldots T)) = \prod_t \sum_{n(t)=0}^{\infty} P(F(t), n(t)|\lambda(t, S(t), \theta), F(0 \ldots t - \Delta t))$$
$$= \prod_t \sum_{n(t)=0}^{\infty} P(n(t)|\lambda(t, S(t), \theta)) P(F(t)|n(t), F(t - \Delta t))$$

The expression on the first line relies on the instantaneous dependency on the rate (dependencies on stimulus history are mediated through the rate), and an explicit integration (marginalization) over the unobserved spike counts, $n(t)$. The second line stems from our assumptions that the fluorescence at time $t$ is independent of the spiking and fluorescence history given the calcium level at the previous time interval (assumed to decay exponentially). Since we cannot directly measure the calcium level we approximate it with the fluorescence at the previous time interval. The firing rate that we wish to infer is notated $\lambda(t, S(t), \theta)$, allowing for an explicit dependency on time, stimulus or behavioral response, and/or tuning parameters. We assume $P(n(t)|\lambda(t, S(t), \theta))$, the distribution of spike counts given the rate, is Poisson,

Since we assume exponential decay of calcium levels and Gaussian measurement noise, the fluorescence at time $t$ given the spike count and the fluorescence at the previous time step is normally distributed with an expected value $\mu(t) = \alpha c(t - \Delta t) \exp\left(-\frac{\Delta t}{\tau}\right) + a \cdot n(t)$ and variance (due to measurement noise) $\sigma^2$, where $c(t)$ represents the intracellular calcium level, $\alpha$ is a scaling factor, $\tau$ is the decay time constant of the calcium signal, $a$ is the increase in calcium level caused by a single spike. Since calcium levels are not observed in the experiment, we approximate the scaled calcium level in the previous time step with the fluorescence level in the previous time step. Thus $P(F(t)|n(t), F(t - \Delta t))$ is Normal with variance $\sigma^2$ and mean $F(t - \Delta t) \exp\left(-\frac{\Delta t}{\tau}\right) + a \cdot n(t)$. Although the Markov assumption on which we rely is also the critical assumption underlying well-known sequential estimation procedures such as the Kalman filter, note that our solution is described in terms of the previously measured fluorescence and not the previously estimated calcium level. If we assume that the initial calcium level has a Gaussian prior, then due to the sum in eq. (1), the posterior would be a sum of increasingly many Gaussians in each step. This is computationally prohibitive, and thus we chose to use the measured fluorescence, which provides an unbiased and cost-free estimate (given our assumptions) of calcium level (up to scaling).

In order to maintain convexity of the likelihood, the parameters $\sigma^2, a, \tau$ must be estimated independently from the data (we used basis pursuit to do so, as described below).

Since we cannot compute the infinite sum in eq. (1), we can either approximate the Poisson spike count distribution with an exponential distribution (changing the sum over spike counts to an integral e.g. Vogelstein et al. 2010), or truncate the sum at some maximal value $n_{max}$, ignoring the contribution of greater spike counts. We find that the latter approach yields substantially more accurate results for the frame rate of most imaging experiments. In particular, spike counts in a single imaging frame of duration ~100 ms are bounded and tend to be low, and thus truncation does not result in a substantial loss in accuracy (see Appendix for more details). We define the log-likelihood of the model as

(2)
$$L(F(0 \ldots T); \theta, S(0 \ldots T)) = \sum_t \log \left( \sum_{n(t)=0}^{n_{max}} P(n(t)|\lambda(t, S(t), \theta)) P(F(t)|n(t), F(t - \Delta t)) \right)$$

For sequential estimation, we infer a rate for each point in time, namely $\lambda(t, S(t), \theta)) = \theta_t$. For direct non-parametric estimation we fit a rate to each one of a discrete set of stimuli, i.e. $\lambda(t, S(t), \theta) = \theta_{S(t-\Delta t)}$. And for direct parametric estimation of Von Mises tuning curves the rate was defined as $\lambda(t, S(t), \theta) = c \cdot e^{k \cdot \cos(S(t-\Delta t) - \mu)}$ and $\theta = (c, k, \mu)$. In all cases an additional baseline term was included in the model to account for periods where no stimulus was presented. To impose a positive rate without introducing constraints, the parameters defined the logarithm of the rate instead of the rate itself, for example – $\log \lambda(t, S(t), \theta) = c' + k \cdot \cos(S(t - \Delta t) - \mu)$.

The inferred parameters were those that maximized the log-likelihood *L*. Optimization was performed using a trust-region algorithm in MATLAB. Note that for sequential estimation and direct non-parametric estimation (not for parametric Von Mises tuning), the likelihood is convex in the parameters and therefore the solution is a global maximum. See Appendix for more details.

To implement a smoothness prior we added a term penalizing the sum of squared differences between neighboring pixels in the rate map (8 neighbors per pixel; toroidal boundary conditions) to eq. (2). We implemented this prior in the relevant section of Results. For all other sections no smoothing or post processing was performed.

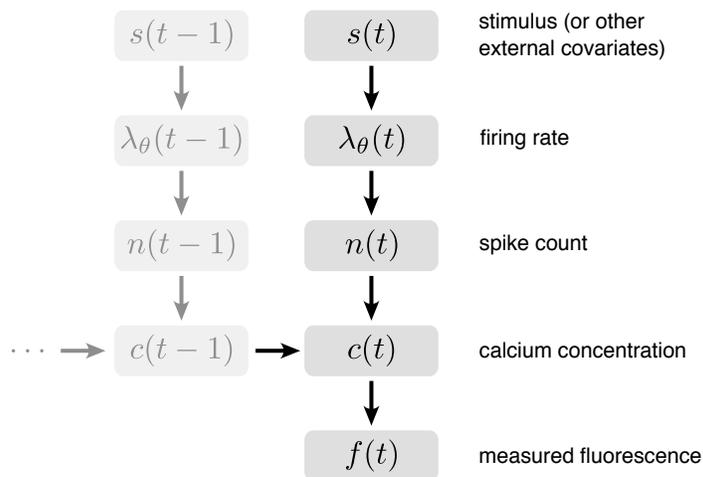

**Figure 1**. Graphical representation of model assumptions and dependencies. The only observable variables are the stimulus *s(t)* and the fluorescence *f(t)* , the rest are hidden. All dependencies are instantaneous and simultaneous in time, except calcium levels which depend on values at the previous time step (through exponential decay), and (potentially) the stimulus history.

## *Surrogate data*

Surrogate data were generated following the scheme in fig. 1: (1) A rate was chosen, either by sampling from a Gaussian distribution, or as a deterministic (tuning curve) function of a time-varying stimulus, $\lambda(t) = f(\theta, S(t))$, where $\theta$ are the parameters defining the tuning, and $S(t)$ represents the stimulus. $S(t)$ may be multidimensional and can include history up to (but not including) time *t*. When neural coupling was inferred, we simply replaced the stimulus with the fluorescence of the coupled neuron, i.e. $\lambda(t) = f(\theta, F_2(t))$. $F_2(t)$ represents the fluorescence of the coupled neuron, and may include history up to (but not including) time *t*. (2) Spike counts were generated by sampling from a Poisson distribution with the given rate $n(t) \sim Poi(\lambda(t))$. (3) A calcium level was generated by convolving the spike count with an exponentially decaying filter, (i.e. spikes in each time bin caused an instantaneous increase in calcium proportional to the spike count, followed by an exponential decay). (4) Finally, the fluorescence trace was generated by adding Gaussian measurement noise. Unless noted otherwise, 50 randomly generated repeated trials were used in each experiment.

## *Linear Model*

For comparison we also fit the data with a linear model, $F = k * (A \cdot S)$, where $F$ is the fluorescence trace, $A$ is a coefficient matrix (containing tuning parameters), $S$ is the stimulus matrix, and $k$ is the exponential decay filter that captures the calcium dynamics. This model predicts a fluorescence level for each stimulus value. The equation is linear in the tuning curve (contained in A), which can thus be estimated using conventional (least-squares) regression.

## *Experimental Procedures*

All experimental procedures were conducted in accordance with the UK Animals Scientific Procedures Act (1986). Experiments were performed at University College London under personal and project licenses released by the Home Office following appropriate ethics review.

## *Surgical procedures and expression of calcium indicator*

Experiments were performed either in Camk2a-tTA; EMX1-Cre ; Ai93(TITL-GCamp6f) triple transgenic mice (Madisen et al. 2015), expressing calcium indicator GCaMP6f in all the cortical Camk2a-positive excitatory neurons, or in wild type C57BL6/j mice where GCaMP6f was expressed in all neurons of a local region using a virus injection.

Using aseptic techniques, mice were implanted with a cranial window over the right visual cortex as previously described (Andermann, Kerlin, and Reid 2010; Andermann et al. 2011). An analgesic (Rimadyl, 5 mg/Kg, SC) was administered on the day of the surgery and in subsequent days, as needed. Dexamethasone (0.5 mg/kg, IM) was administered 30 min prior to the surgery to prevent brain edema. The animal was anesthetized with Isoflurane (1-2% in 100% Oxygen), body temperature was monitored and kept at 37-38 °C using a closed-loop heating pad, and the eyes were protected with ophthalmic gel (Viscotears Liquid Gel, Alcon Inc.). The head was shaved and disinfected, the cranium was exposed and covered with biocompatible cyanoacrylate glue (Vetbond). A stainless steel head plate with a 7 mm round opening was secured over the skull using dental cement (Super-Bond C&B,

Sun Medical Co. Ltd., Japan). Then, a 3-4 mm craniotomy was opened over the visual cortex (centered at -3.3mm AP, 2.8 ML from bregma). Finally, the craniotomy was sealed with a glass cranial window, attached to the skull using cyanoacrylate glue and dental cement. The window was assembled from a 5 mm outer round cover glass cured to 1-2 smaller inserts (3 mm, Warner Instruments, #1 thickness) with index-matched UV curing adhesive (Norland #61). The animal was allowed to recover for at least 4 days before further experimental procedures.

For the wild type animals, before sealing the craniotomy we injected an AAV1.Syn.GCaMP6f.WPRE.SV40 virus (100 nl, titer of 2.4e12 GC/ml, UPenn Vector Core) 250-300 μm below the V1 surface. Virus reached expression level suitable for imaging ~2 weeks after the injection.

## *Two-photon calcium imaging*

For the imaging experiments mice were head-fixed under a resonant-scanning two-photon microscope (B-Scope, Thorlabs). Mice were free to run on an airflow-suspended spherical treadmill during the imaging sessions. The microscope was controlled using ScanImage v4.2 (Pologruto, Sabatini, and Svoboda 2003). A low magnification (x16) high NA (0.8) water immersion objective lens (Nikon) was mounted on a piezoelectric z-drive (PIFOC P-725.4CA, Physik Instrumente) allowing multi-plane imaging. Excitation light (970 nm, 30-60 mW at the sample) was provided by a femtosecond laser (Chameleon Ultra II, Coherent). Images (512x512 pixels, with field of view of 340-500 microns) were acquired at 30 Hz. This high imaging rate was temporally divided into 3-5 different depths spaced 50-60 microns apart, resulting in an acquisition rate of 6-10 Hz per imaging plane.

## *Visual stimulation*

Visual stimuli were generated in Matlab (MathWorks) using the Psychophysics Toolbox (Brainard 1997; Kleiner et al. 2007) and displayed on 3 gamma-corrected LCD monitors (refresh rate 60Hz) arranged at 90 degrees to each other. The mouse was positioned at the center of this U-shaped arrangement at the distance of 20 cm from all three monitors so that the monitors spanned ±135 degrees of horizontal and ±35 degrees of the vertical visual field of the mouse. For rate mapping experiments we used sparse spatial white noise stimuli. Patterns of sparse black and white squares (4.5-7.5 degrees of visual field) on gray background were presented at 5 Hz. The probability of each square to be not gray was 2-5% and independent of other squares. For orientation tuning experiments we presented 0.5 s long drifting gratings (size 60 degrees, contrast 50%, spatial frequency 0.05 cpd, temporal frequency 2 Hz, 4 different phases). The position of the grating was selected to match the retinotopic location of the imaged region.

## *Data preprocessing*

We removed the baseline from fluorescence traces using robust local regression estimation (Ruckstuhl et al. 2001). This procedure also provides an estimate of the measurement noise standard deviation $\sigma$. The spikes were estimated using basis pursuit denoising (van den Berg and Friedlander 2008) and the magnitude, $a$, was set to be the 95$^{th}$ percentile of the coefficient values in the solution of the sparse inverse problem. Calcium decay times, $\tau$, were generally set to 0.5 s. All aforementioned values were inspected manually and adjusted if necessary.

## Results

Given fluorescence measurements, and stimuli, behavior, or other external covariates, we estimate rates directly by maximizing the likelihood given in Eq. (2) (see Methods). Unlike deconvolution approaches, which aim to explicitly infer spikes, our approach integrates over the unobserved spikes, directly inferring the underlying rate.

### *To deconvolve or not to deconvolve?*

As an illustrative example, consider the ubiquitous task of estimating the mean firing rate across repeated trials. The intuitive approach would be to first estimate the spike count, in each trial, and then average these counts across repeated trials to obtain rates. We refer to this approach "sequential estimation". Alternatively, our method aims to estimate the rate directly from the calcium measurements across all repeated trials. We refer to this as "direct estimation".

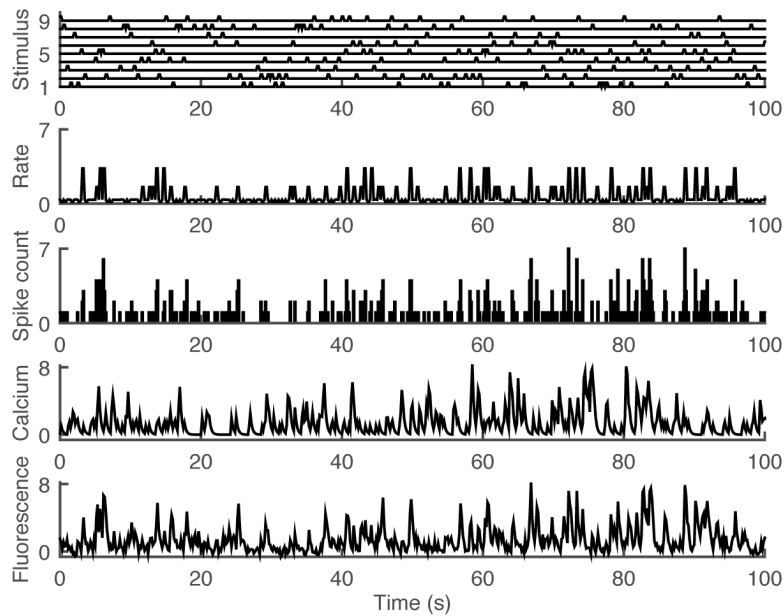

**Figure 2.** A Simulated example of temporal signals in the fluorescence generation process. At each point in time at most one of 9 discrete stimuli is 'on' (1st row). Each stimulus is deterministically associated with a rate (2nd row). Spike counts are samples from a Poisson process with the given rate (3rd row). Calcium levels are a convolution of the spike counts with an exponential filter – i.e. each spike causes an instantaneous rise in calcium followed by an exponential decay (4th row). Finally, the measured fluorescence is a scaled version of the calcium, corrupted with Gaussian measurement

noise (5th row). The goal is to estimate the relationship between stimulus and rate, given the observed fluorescence signal.

To compare the two approaches, we generated data as illustrated in Fig. 2. Fifty noisy repeats with the same underlying rate were randomly generated. The log of the rate for each time bin was drawn from a Normal distribution ($\mu = -4, \sigma = 1.5$). We then estimated the rate in one of three ways: (a) By fitting eq. (2) from Methods simultaneously to all 50 repeats; (b) By fitting eq. (2) to each individual repeat and averaging across repeats; and (c) By using the deconvolution method of (Vogelstein et al. 2010) to estimate spike count/rate on each repeat, and averaging these across repeats. As can be seen in fig. 3A, fitting all repeats simultaneously substantially improves the estimate of the rate. Moreover, estimates based on averaging single trials do not converge to the true value of the rate. This occurs because low rates are systematically overestimated – for very low (or zero) spike counts, the effects of measurement noise are asymmetric, generally leading to an increase in the number of estimated spikes, since the number of spikes must remain non-negative.

We also note that even if the goal of the analysis is to estimate the spike count in each time bin, our approach provides a more accurate estimate of the spike count than the deconvolution approach (fig. 3B). This is a consequence of the fact that we explicitly model a time varying rate, and assume that spike counts follow a Poisson distribution (which is the true distribution for the simulated data). Therefore, for the remainder of this article, we will use the rate estimates from eq. (2) averaged across trials to perform sequential estimation, rather than averaging spike count estimates provided by deconvolution.

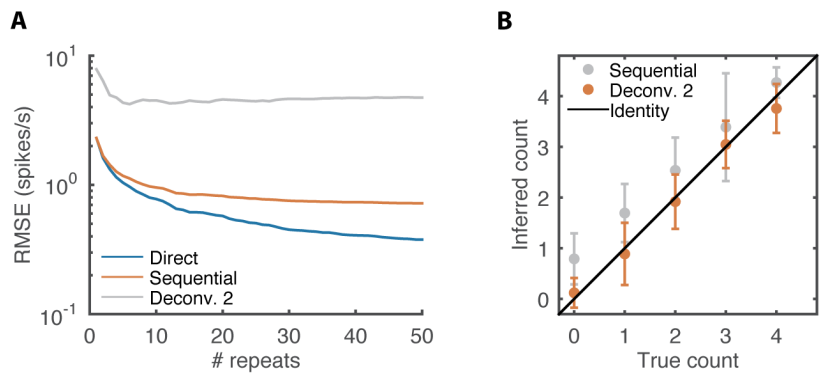

**Figure 3**. Estimating rates and spike counts from simulated calcium fluorescence data. **A**. 50 fluorescence traces with the same underlying rate were randomly generated (measurement noise

$\sigma = 0.5$). The rate or spike count in each repeat was estimated using eq. (2) (red) or a deconvolution algorithm (yellow; Vogelstein et al. 2010), respectively, and then averaged across repeats (sequential estimation). Alternatively, the rate was simultaneously fit to all repeats using eq. (2) (blue; direct estimation). The root mean squared error, relative to the true underlying rate used to generate the data, is shown as a function of the number of repeats used in the estimation process. Though unintuitive at first, estimating the spike counts and then averaging is not the same as estimating the average count directly. The latter proves to be more accurate. **B**. Estimating spike counts from fluorescence. We used eq. (2) to estimate the rate, or the deconvolution algorithm of Vogelstein et al. 2010 to estimate the spike count/rate at each point in time. Our estimate of the rate (red; error bars represent standard deviation across inferred rates) proves a better estimate of the true spike count than that provided by the deconvolution algorithm (gray).

## *Estimating tuning curves*

One of the shortcomings of deconvolution approaches is that they are oblivious to the experimental structure. That is, they do not take into account which stimulus, behavior, or experimental condition was present at each point in time. While subsequent stages of analysis may take the experimental structure into account, they cannot correct errors made during the deconvolution stage. Since our method assumes that the stimulus or experimental condition influences the rate directly, rather than the spike count (which arises from the rate) it is able to leverage the structure of the experiment, yielding better estimates.

To quantify the advantage of the direct approach over sequential estimation, we generated data from the generative model described in fig. 2, where the rate was determined by the stimulus condition at the previous time step. At each point in time there was at most one of 9 different stimuli present. We estimated a tuning curve by either estimating the count at each point in time and then averaging these counts over each of the 9 stimuli (sequential estimation), or by directly fitting a stimulus dependent rate to all data simultaneously (direct estimation).

When measurement noise levels were low, the deconvolution process produces an accurate estimate of spike counts, and both estimates are accurate. But at higher noise levels, the deconvolution process becomes substantially less accurate, and, consequently, the sequential estimation of firing rates becomes less accurate (fig. 4A,B). As an additional comparison, we also show estimates derived from linear regression (see methods). While linear estimates are less sensitive to noise they are substantially less accurate than either sequential or direct estimation.

Robustness to noise is of particular importance in imaging experiments, which are often designed for high throughput simultaneous measurements from many cells. The field of view of a typical imaging experiment may easily contain thousands of neurons, and while some may be relatively 'clean' many are noisy. To harness the potential of calcium imaging one would like to 'dig' as deeply into the data as possible. To this end, we note that the direct estimation method requires fewer samples (shorter experiment) to achieve the same accuracy as sequential estimation (fig. 4C). This is of particular importance for imaging experiments carried out in awake head-fixed animals in which animal welfare considerations dictate strict limits on experimental duration.

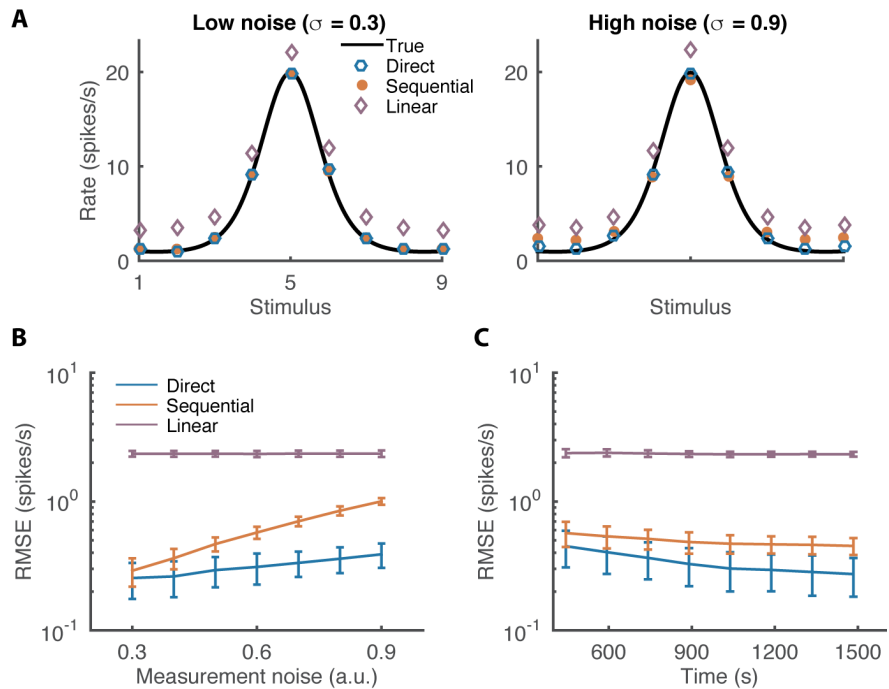

**Figure 4**. Estimating tuning curves from simulated data. **A**. Fluorescence traces were generated following the generative model (fig. 1), with stimulus-driven rates determined by a tuning curve (shown in black), for two levels of measurement noise $\sigma$. The rate associated with each stimulus was estimated in three ways: (1) By first estimating the rate at each point in time and then averaging the rates following each stimulus (sequential estimation; red crosses); (2) By fitting a stimulus dependent rate directly to the data (direct estimation; blue circles); and (3) Using linear regression (yellow triangles). **B**. Root mean squared error of the estimated tuning curve is plotted as a function of measurement noise. Error bars represent standard deviations across 50 randomly generated fluorescence traces. As measurement noise increases the advantage of direct estimation becomes more pronounced. **C**. Estimation error as a function of experiment duration (color legend as in panel B). Direct estimation requires less data to achieve the same accuracy.

## *Application to data*

Thus far, we've compared the performance of the sequential and direct approaches applied to simulated data, which obeys all of the model assumptions. Real neural data, on the other hand, is unlikely to obey these assumptions. In the following section we apply our method to calcium imaging data obtained from mouse visual cortex.

### *Recovering firing rates*

As before, we begin with the basic task of estimating firing rates from imaging data. We used a dataset described in (Chen et al. 2013), and publicly available on CRCNS.org. This dataset includes simultaneous imaging and loose-seal cell-attached electrical recordings, under repeated visual stimulation. The electrical recordings provide a ground-truth measurement of the true spike counts.

The data set consists of 11 cells recorded for 6 trials of 40 s each (multiple recordings per cell). During each trial a visual stimulus was presented for 2 s. We estimated firing rates from the imaging data either directly or sequentially (example in fig. 5A). Sequential estimation performed particularly poorly in low firing rate, substantially overestimating the baseline firing rate. For almost all cells, direct estimation outperformed sequential estimation by roughly a factor of two (fig. 5B).

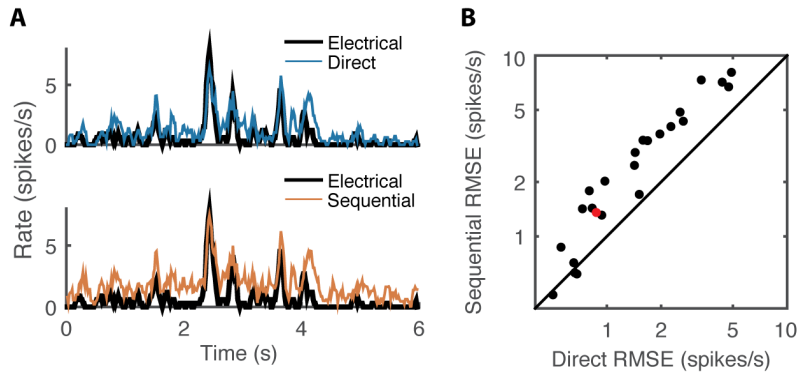

Figure 5. Estimating firing rates from simultaneous imaging and electrical recordings. **A**. The mean firing rate across repeats for an example cell was estimated either by averaging electrically recorded spike counts across 40 repeated trials (black), by averaging spike counts inferred from individual trials of imaging data (bottom; red; sequential), or by directly estimating the rate across trials from imaging data (top; blue; direct). **B**. Root mean squared error (relative to the rate estimated from electrical recordings) of the sequential method is plotted against that of the direct method (black line indicates equality). Each point corresponds to estimates for a single cell, obtained on 40 repeats of a 6s stimulus (n=24 recordings). Red point corresponds to the example in panel A.

### *Estimating rate maps*

Next, we applied our approach to estimate the rate maps of neurons in mouse visual cortex. Mice were head-fixed on a floating ball and viewed sparse noise stimuli on a screen while calcium activity (reported by GCaMP6f) was recorded using a two-photon microscope (see Methods for details). The stimulus was a 10x36 square pixel grid, where each pixel had a 2.5% chance of being either white or black and a 95% chance of being gray. The goal of the analysis was to identify the region in the visual field in which changes to the light affect the neuron's response.

As in the previous sections, we estimated the rate associated with a pixel either by first estimating the rate in each time bin, and then averaging all time bins for which that pixel was black/white (sequential estimation), or by directly fitting a pixel-intensity dependent rate to the data (direct estimation).

Figure 6A shows the inferred rate maps for an example neuron. This corresponds to each model's expected rate given each pixel being black, independent of other pixels. Since the stimulus is sparse (see methods) interactions between pixels are negligible. Visual inspection reveals that the sequential approach produces a noisier estimate of the rate map. We quantified the noise level in the rate map using the median absolute deviation (MAD). We chose the MAD over the standard deviation as a measure of variability/noisiness, because it is less sensitive to outliers arising at the extrema of the rate maps. For all cells in this data set exhibiting clear rate map structure, fitting the data directly with a position-dependent rate resulted in less noisy rate map estimates compared to a sequential estimate or linear regression (fig. 6B).

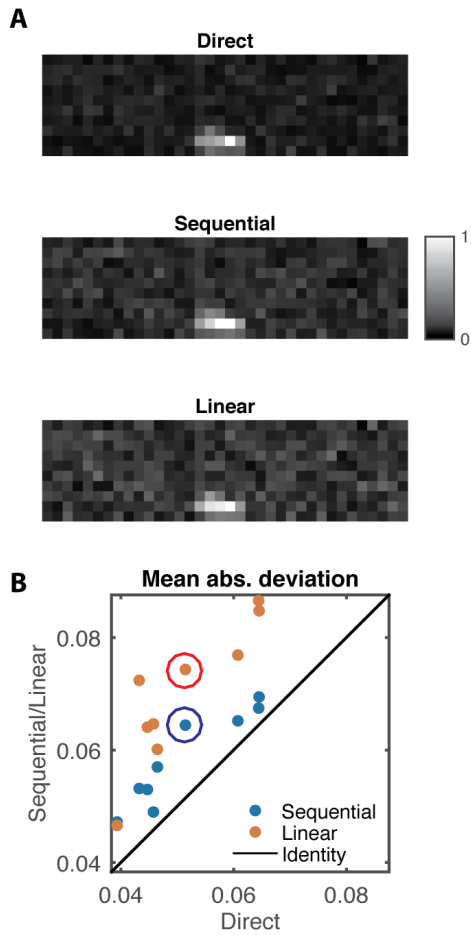

**Figure 6**. Estimating rate maps from data. **A**. Rate maps of mouse visual cortical neurons were estimated using direct estimation (top), sequential estimation (middle) and linear regression (bottom; see text for details). The intensity of each pixel represents the rate (rescaled) associated with that pixel being black (this is an off cell). The direct estimation approach provides the least noisy estimate of the rate map (no smoothing or regularization was applied). **B**. The quality of the rate map estimate was quantified using the median absolute deviation (MAD) of the entire rate map. The MAD for the sequential estimates (red) and linear regression (yellow) are plotted against the MAD for the direct estimates (n = 9 neurons; circled data points correspond to the example in panel A). For all cells the direct estimate proved least noisy.

Above we measured the 'goodness' of the rate maps by measuring their smoothness. Our framework is sufficiently general that we can explicitly incorporate this prior knowledge about the smoothness of rate maps into the model. This can be done by adding a penalty proportional to the sum of squared differences between neighboring pixels to the log likelihood. In fig. 7 we show the impact of adding a smoothness prior on rate map estimation. Note that here for clarity we show the linear portion of our model, before the nonlinearity, and thus a log-rate map.

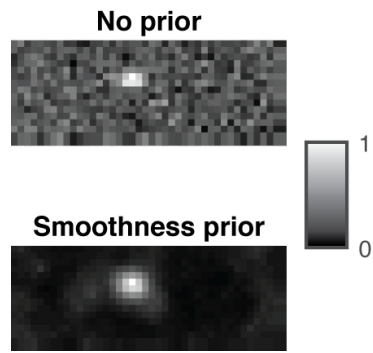

Figure 7. Estimating rate maps with a prior preference for smoothness. **A**. The linear spatial filter portion of the model estimated for an example neuron from responses to a sparse noise stimulus. **B**. The same as panel A, but with the smoothness prior imposed.

### *Estimating parametric orientation tuning*

We now turn to the problem of estimating orientation tuning curves from responses to drifting grating stimuli (same neurons as in previous section; see Methods). Orientation tuning curves are often summarized by a scaled Von Mises probability density function (Swindale 1998), which is defined by only three parameters – mean, width and scale. The mean of the Von Mises fit is the preferred orientation of the neuron.

This widely used parametric form of the tuning curve allows us to highlight another feature of our model. Since we explicitly model the relationship between the stimulus and the response, the parametric form may be incorporated into the objective function, allowing us to directly estimate the parameters (note that the parametric form can impact the convexity of the likelihood, and thus the complexity of the optimization problem; see appendix for more details). For comparison, we also consider a sequential scheme, in which we first estimate rates on individual trials, then average them, and finally fit a Von Mises function to these averages.

To assess the quality of the fitted tuning curves, we measured the variability of the estimate across randomly selected subsets of the data. A better estimator should exhibit less variability across the different subsets.

With limited data (less than 5 minutes) estimating the parameters defining a Von Mises tuning curve directly from the data proves more reliable (fig. 8) compared to sequential estimation. This is because we are only fitting 3 parameters to the entire trace, and we are taking into consideration any uncertainty that we have in the estimates of the rate. In the sequential approach, each stage is oblivious to any uncertainties in the previous stage. As the amount of data increases, both approaches perform consistently, but this is partially because the overlap between the randomly chosen subsets necessarily increases.

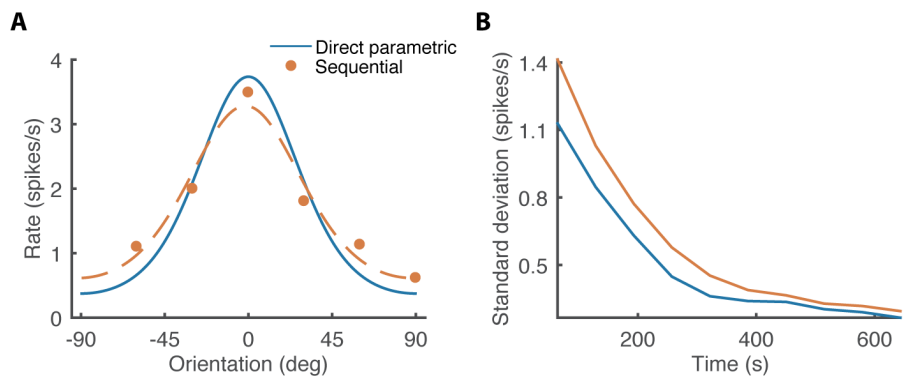

**Figure 8**. Estimating orientation tuning curves from data. **A**. Orientation tuning curves of a mouse visual cortical neuron was estimated either sequentially (estimating rates and then averaging the rates associated with each orientation; red circles), or by directly fitting a parametrically defined Von

Mises shaped tuning curve directly to the fluorescence data (blue curve). **B**. Subsets of data of different lengths were randomly selected and the tuning of the neuron was estimated. The standard deviation of these estimates is plotted as a function of the length of the subset used for the estimation. The direct parametric approach provides a more reliable estimate of the tuning for any given amount of data.

### *Inferring neural coupling*

Previous work has demonstrated the value and importance of modeling the coupling between neurons (Pillow et al. 2008; Gerhard et al. 2013; Okatan, Wilson, and Brown 2005). Neural coupling is predominantly mediated through spikes and the ensuing synaptic events. Our direct inference approach, however, does not provide an estimate of the spikes. Is it still possible to infer neural coupling?

Though actual spike times are never inferred throughout our analysis, the observed fluorescence trace provides a noisy filtered version of the spikes. Therefore we investigated whether coupling between neurons, at the level of spikes, can be inferred using the fluorescence signal. Note that this analysis can only infer interactions occurring at the relatively slow sampling rate of the experiment (~10 Hz), and not those occurring at the time scale of synaptic transmission (>100 Hz).

To test whether neural coupling can be inferred from fluorescence, we simulated two neurons, in which the spikes of neuron 2 multiplicatively influence the rate of neuron 1 through a predefined temporal filter, $c_s$. We then inferred the rate of neuron 1 from its fluorescence trace, using the fluorescence trace of neuron 2 as a covariate (as if it were an external stimulus). This procedure provides us with a maximum likelihood estimate of neuron 1's rate and an estimate of the coupling between the neurons. We will denote this inferred filter $c_f$. One must keep in mind that $c_s$ and $c_f$ are not expected to be identical, as the former operates on spikes while the latter operates directly on the fluorescence signal. Yet, it is easy to see that if we want to generate a filter which will mimic $c_f$'s output when applied to spikes, we simply need to convolve $c_f$ with an exponential filter, the same convolution process that converts spike trains to calcium traces. We denote this filter $\hat{c}_s$.

As expected, while the inferred filter $c_f$ does not match the coupling filter used to generate the data, once we convolve it with an exponential decay to obtain $\hat{c}_s$ produces a very good estimate of the true coupling between the neurons (fig. 9). This suggests that our approach enables inference of spike based coupling between neurons, although spikes are never inferred in the process.

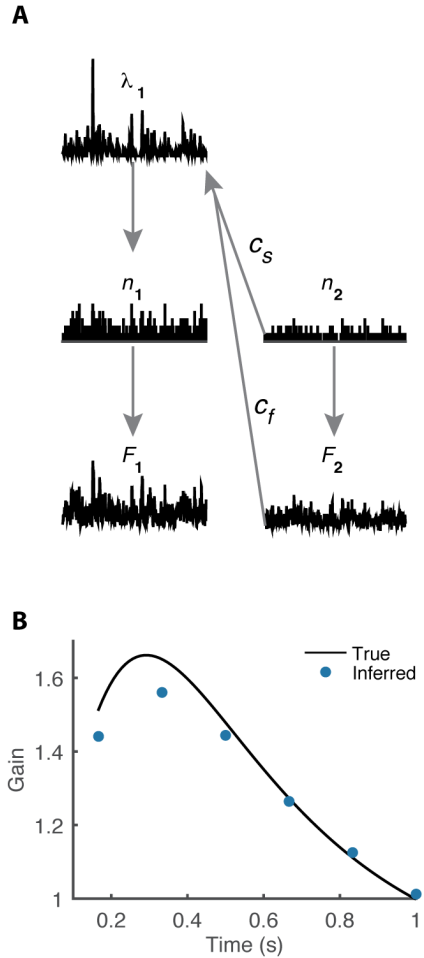

**Figure 9**. Inferring spike based coupling from fluorescence. **A**. The rate of neuron 1 (top) is directly influenced by the spiking activity of neuron 2 through a coupling filter $c_s$, yet our task is to infer this coupling by observing fluorescence alone. **B**. We generated surrogate fluorescence data using a known coupling filter $c_s$ (black line). We then estimated the neural coupling $c_f$ by inferring the rate of neuron 1 using the fluorescence of neuron 2 as a covariate (a fluorescence dependent rate). The inferred coupling $c_f$ was convolved with an exponential decay to obtain $\hat{c}_s$ (blue circles; see text for details).

## *Discussion*

We have shown that sequential estimation of neural firing rates from calcium imaging data, by first estimating spike counts and then averaging these to obtain firing rates and/or tuning curves, can produce strongly biased results. As an alternative, we've developed a direct method that operates by integrating over the unobserved spike counts. Our analysis shows that direct estimation outperforms sequential estimation in simulation, as well as on imaging data. The method is flexible, allowing estimation of firing rate dependencies on any measured covariate (e.g., movement direction, location in space, task condition, behavioral state), using calcium imaging obtained from either repeated trials, and/or by assuming a parametric relationship between covariates and firing rates.

### *Calcium trace deconvolution*

The term "deconvolution" stems from the assumption that intracellular calcium levels can be described as a convolution of the spike train with a linear filter. Deconvolution provides a natural preprocessing stage for the analysis of imaging data, extracting estimates of the underlying spikes, after which standard methods may be used to analyze the spike counts. Deconvolution algorithms are based on a variety of different methodologies including Bayesian inference (Vogelstein et al. 2010; Theis et al. 2015), basis pursuit (Grewe et al. 2010), matching pursuit (Dyer et al. 2010), among others (Oñativia, Schultz, and Dragotti 2013). The method of (Vogelstein et al. 2010), in particular, is based on the same generative model we have used in this article. But our direct method differs in that: (a) it explicitly includes a stimulus-dependent, time varying rate, whereas in (Vogelstein et al., 2010) the rate is fixed and is used as a sparse prior on spike counts; (b) it integrates over the unobserved spike counts, rather than attempting to explicitly estimate them; (c) it assumes a Poisson spiking process, rather than substituting a more computationally tractable exponential distribution.

These differences, while seemingly minor, result in substantial improvements in the accuracy of rates inferred from calcium data (see fig. 3). In particular, the presence of noise in the imaging process ensures that the deconvolution results will be imperfect, but subsequent analyses generally ignore the uncertainty associated with the spike count estimates. As we have shown this can lead to biases when estimating firing rates and tuning curves from the data, especially for low firing rates. By integrating out the spike counts we are able to take the uncertainty associated with spike count estimation into account, and consequently provide a procedure that is more tolerant to noise. Furthermore, the direct approach proves more statistically efficient, which in practice can translate into shorter experiments.

Similar arguments in favor of direct estimation have been made in the past with regard to spike sorting. Namely it has been shown that it may be preferable estimate tuning curves directly from raw voltage traces, rather than to first sort the spikes and then fit the tuning curves (Ventura 2009b; Ventura 2009a). Recently it was shown that an animal's position in space can be decoded with better accuracy directly from hippocampal spike waveforms, than by a sequential scheme in which spikes are first sorted and then used for position estimation (Kloosterman et al. 2014; Deng et al. 2015).

For both spike sorting and deconvolution it is important to remember that the output of any algorithm is only an estimate of spike times/counts, and should be treated as such. Subsequent analyses that ignore the uncertainty in these estimates may accumulate errors and biases. This is not a problem when the signal to noise ratio is very large and spikes can be recovered with high accuracy, yet these cases are increasingly rare as high-throughput physiological recordings are becoming more prevalent.

## *Model assumptions*

The elements of our model are commonplace in the literature, and are not unique to our approach, such as Poisson spiking (Dayan and Abbott 2001) or Gaussian measurement noise (Wilt, Fitzgerald, and Schnitzer 2013).

An additional assumption, central to most deconvolution approaches, is that calcium levels are the result of a linear convolution of the spike train with an exponential decay filter. In general, peak calcium levels are not directly proportional to spike count, and decay time can vary with spike count (Akerboom et al. 2012; Badura et al. 2014; Chen et al. 2013). These observations may be explained by non-linear summation of spike 'signatures' (Nauhaus, Nielsen, and Callaway 2012), which is most pronounced for high firing rates. It is not immediately clear how to incorporate a nonlinearity into the convolution process in our generative model while maintaining computational tractability, thus it may be advisable to remain near the linear operating regime of the indicator. Further we assume that the spike

'signature' includes an instantaneous rise. Although this is an idealization, for imaging experiments with the latest calcium indicators the rise time is typically on the order of the sampling rate. Thus we do not expect this assumption to greatly affect the model.

### *Region of interest*

A seemingly independent problem when analyzing imaging data is that of Region Of Interest (ROI) detection (Mukamel, Nimmerjahn, and Schnitzer 2009). Recently several groups have noted that the ROI detection and deconvolution problems are in fact related. Knowledge of spike times can clearly guide ROI detection and vice versa. Consequently some authors have explored the combination of these two analysis steps into a single optimization problem, trying to solve for the ROIs and spike times/counts simultaneously (Diego Andilla and Hamprecht 2014; Pnevmatikakis et al. 2014). Following our insights from the current work, we suggest that when the goal of the analysis is rate (or tuning curve) estimation, it may prove beneficial to estimate the rate directly from the raw images. This may prove to be a computationally prohibitive task and we defer this for future research.

### *Conclusion*

As imaging methods become an increasingly popular way to measure brain activity, it is important to develop algorithms to analyze these data. We chose to focus on experiments in which the goal of the analysis is a firing rate, an average spike count histogram, or a tuning curve. Although this does not encompass all possible analyses, it is sufficiently broad to include a wide range of experiments.

We argue that the 'correct' approach to this estimation problem is to integrate out the unobserved spikes to obtain the quantities of interest. Analogous proposals have been made in the context of electrophysiological data (Ventura 2009b; Ventura 2009a; Kloosterman et al. 2014; Deng et al. 2015). Properly integrating over "nuisance variables" is a well-known theme in the statistical estimation and machine learning communities, but is notorious for being computationally expensive. Our direct method of estimating firing rates and tuning curves from imaging data provides state-of-the-art estimates of rates by integrating out the spike counts, yet is efficient enough to run on a standard computer.

## *Acknowledgements*

This work was supported by the James S. McDonnell Foundation (EG), the Wellcome Trust (LFR, MK, and MC), and the Howard Hughes Medical Institute (EPS). MC holds the GlaxoSmithKline / Fight for Sight Chair in Visual Neuroscience.

# *Appendix*

## *Properties of the model likelihood*

As shown in the methods, the log-likelihood of our model can be written as

(A.1)
$$L(F(0 \ldots T); \theta, S(0 \ldots T)) = \sum_t \log \left( \sum_{n(t)=0}^{n_{max}} P(n(t)|\lambda(\theta, S(0 \ldots t))) P(F(t)|n(t), F(t - \Delta t)) \right).$$

Our task is to find the set of tuning parameters $\theta$ that maximize the likelihood. These parameters may explicitly represent the rate in response to different stimuli or define a parametric form like the Von Mises function.

Since the rate must be non-negative it is convenient to work with the log-rate. We will express the derivative in terms of the derivative of the log-rate, which may depend on time, stimulus, or be a parametric function of the stimulus. We use the notation $\lambda(\theta, t)$ to mean either $\lambda(t)$ when are estimating rate directly or $\lambda(\theta, S(t))$ when the rate depends on the stimulus through parameters $\theta$. For brevity we denote the complete fluorescence trace and stimulus by $F$ and $S$ respectively.

Incorporating the Poisson spiking assumption, the log likelihood is equal to

$$L(F; \theta, S) = \sum_t \log \left[ \sum_{n=0}^{n_{max}} \frac{\lambda(\theta, t)^n e^{-\lambda(\theta, t)}}{n!} \cdot P(F(t)|n(t), F(t - \Delta t)) \right]$$

The gradient of the likelihood with respect to the $i^{th}$ parameter can be written as

$$\frac{\partial \log L(F; \theta, S)}{\partial \theta_i} = \sum_t \frac{\partial \log \lambda(\theta, t)}{\partial \theta_i} \left( \frac{\sum_{n=0}^{n_{max}} n \frac{\lambda(\theta, t)^n}{n!} P(F(t)|n(t), F(t - \Delta t))}{\sum_{n=0}^{n_{max}} \frac{\lambda(\theta, t)^n}{n!} P(F(t)|n(t), F(t - \Delta t))} - \lambda(\theta, t) \right)$$

Multiplying the numerator and denominator by $e^{-\lambda(\theta,t)}$ this can be written as –

$$\frac{\partial L(F; \theta, S)}{\partial \theta_i} = \sum_t \frac{\partial \log \lambda(\theta, t)}{\partial \theta_i} \left( \frac{\sum_{n=0}^{n_{max}} n P(n(t)|\lambda(\theta, t)) P(F(t)|n(t), F(t - \Delta t))}{\sum_{n=0}^{n_{max}} P(n(t)|\lambda(\theta, t)) P(F(t)|n(t), F(t - \Delta t))} - \lambda(t, \theta) \right).$$

The entries of the Hessian are therefore

$$\frac{\partial^2 L(F;\theta,S)}{\partial \theta_i \partial \theta_j} = \sum_t -\frac{\partial \log \lambda(\theta,t)}{\partial \theta_i}\frac{\partial \log \lambda(\theta,t)}{\partial \theta_j}\left[\lambda(\theta,t) - \frac{\sum_{n=0}^{n_{max}} n^2 P(n(t)|\lambda(\theta,t))P(F(t)|n(t),F(t-\Delta t))}{\sum_{n=0}^{n_{max}} P(n(t)|\lambda(\theta,t))P(F(t)|n(t),F(t-\Delta t))} + \right.$$

$$\left.\left(\frac{\sum_{n=0}^{n_{max}} n P(n(t)|\lambda(\theta,t))P(F(t)|n(t),F(t-\Delta t))}{\sum_{n=0}^{n_{max}} P(n(t)|\lambda(\theta,t))P(F(t)|n(t),F(t-\Delta t))}\right)^2\right] +$$

$$\sum_t \frac{\partial^2 \log \lambda(\theta,t)}{\partial \theta_i \partial \theta_j}\left(\frac{\sum_{n=0}^{n_{max}} n P(n(t)|\lambda(\theta,t))P(F(t)|n(t),F(t-\Delta t))}{\sum_{n=0}^{n_{max}} P(n(t)|\lambda(\theta,t))P(F(t)|n(t),F(t-\Delta t))} - \lambda(\theta,t)\right).$$

Note that if $\frac{\partial^2 \log \lambda(\theta,t)}{\partial \theta_i \partial \theta_j} = 0$ (which is the case for sequential estimation or direct non-parametric estimation) then the negative Hessian is positive semi-definite and the negative log-likelihood is thus convex. To see this, let us denote the term in square brackets $k(t)$, $\frac{\partial \log \lambda(\theta,t)}{\partial \theta_i} = q_i(t)$, and $\bar{v}(t) = \sqrt{k(t)}\bar{q}(t)$. Thus we can write $\frac{\partial^2 L}{\partial \theta_i \partial \theta_j} = \sum_t -\bar{v}(t)\bar{v}(t)^T$, i.e. a sum of outer products of a vector with itself. Therefore the negative Hessian is positive semi-definite.

## *Exponential approximation*

As noted in the main text, it is possible to substitute the assumption of Poisson spiking with exponentially distributed 'counts' (these are not truly counts since they are no longer integers). In this case integrating out the spikes can be done more efficiently. Namely, the sum

$$\sum_{n(t)=0}^{\infty} P(F(t),n(t)|\lambda(t),F(0\ldots t))$$

in eq. (1) in Methods is replaced by an integral –

$$\int_0^{\infty} P(F(t),n(t)|\lambda(t),F(0\ldots t))\,dn(t)$$

In addition, now $P(n(t)|\lambda(t)) = \lambda(t)e^{-\lambda(t)n(t)}$.

Though we did not achieve good results with this approximation, we present the equations for the log-likelihood, gradient, and Hessian for completeness. For brevity we will drop the explicit dependence of $\lambda$ on $\theta$.

$$L(F;\theta) = \sum_t [\log \lambda(t) - \lambda(t)\tilde{F}(t) + \log K_1(t)]$$

$$\frac{\partial L(F;\theta)}{\partial \theta_i} = \sum_t \frac{\partial \log \lambda(t)}{\partial \theta_i}\left[1 - \lambda(t)\tilde{F}(t) - \frac{z(t)\sigma\lambda(t)}{K_1(t)} + (\lambda(t)\sigma)^2\right]$$

$$\frac{\partial^2 L(F;\theta)}{\partial \theta_i \theta_j} = \sum_t \frac{\partial \log \lambda(t)}{\partial \theta_i}\frac{\partial \log \lambda(t)}{\partial \theta_j}\left(-\lambda(t)\tilde{F}(t) + 2\sigma\lambda^2(t)\right.$$

$$\left. - \frac{\sigma z(t)\lambda(t)}{K_1(t)}\left[\lambda(t)\tilde{F}(t) + \sigma\lambda(t)\left(\frac{z(t)}{K_1(t)} - \sigma\lambda(t)\right) + 1\right]\right)$$

$$+ \sum_t \frac{\partial^2 \log\lambda(t)}{\partial \theta_i \theta_j}\left(1 - \lambda(t)\tilde{F}(t) - \frac{z(t)\sigma\lambda(t)}{K_1(t)} + (\lambda(t)\sigma)^2\right)$$

Where we define the following intermediate quantities –

$$\tilde{F}(t) = \frac{1}{a}\left(F(t - \Delta t)e^{-\frac{\Delta t}{\tau}} - F(t)\right)$$

$$K_1(t) = \sqrt{\frac{\pi}{2}}\exp\left(\frac{(\lambda(t)\sigma)^2}{2}\right)\cdot \text{erfc}\left(\frac{\lambda(t)\sigma - \frac{\tilde{F}}{\sigma}}{\sqrt{2}}\right)$$

$$z(t) = \exp\left(-\frac{1}{2}\left(\frac{\tilde{F}(t)}{\sigma}\right)^2 + \lambda(t)\tilde{F}(t)\right)$$

### *Implementation considerations*

There are 3 keys to making the task of integrating out the spike count computationally efficient: (1) The assumption that Fluorescence is conditionally independent of the spiking history given the fluorescence at the previous time step and the current spike count. This allows us to avoid a combinatorial explosion, since we only need to sum over all possible spike counts at the current time step alone. (2) The observation that probability of observing a certain fluorescence value at time $t$, given the spike count at time $t$ and the fluorescence at the previous time step, $F(t)|n(t), F(t-\Delta t)$, does not depend on the parameters being optimized. This means that we can calculate this quantity once for different values of $n(t)$ and do not need to reevaluate it during optimization iterations. (3) The truncation of the sum in eq. (1) of Methods, which ignores the contribution of large spike counts. Since the spike count tends to be small for our choice of time bins, this sum can be safely truncated at an appropriate number, which we designate as $n_{max}$.

A proper choice of $n_{max}$ depends on the preparation and the sampling rate. The data analyzed in this study comes from neurons in the superficial layers of mouse visual cortex. Electrophysiology suggests that such neurons rarely spike at rates that exceed ~20Hz (Niell and Stryker 2008). At a sampling rate of 6Hz (as in our data) this means less than 4 spikes per bin. We used a $n_{max}$ value of ~10 spikes/bin, which is equivalent to ~60 spikes/s. To be on the safe side though, we increased the value of $n_{max}$ from 6 to 25 spikes/bin, but did not observe as substantial effect on the rate map estimate (fig. A.1A).

To achieve the best run time $n_{max}$ should be set as low as possible, but as can be seen in fig. A.1B, fitting a 10x36 rate map using our method takes about 0.1 seconds for a wide range of $n_{max}$ values (on a 2014 Apple MacBook Pro laptop).

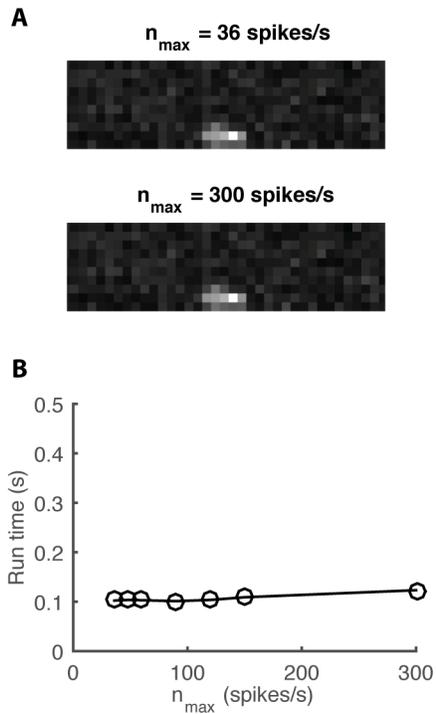

**Figure A.1**. Setting the maximal spike rate and its effect on run time. **A**. The same rate map as in fig. 6 was fit using different values of $n_{max}$. Changing $n_{max}$ from 36Hz to 150 Hz has no visible effect on the inferred rate map. **B**. The time it takes to fit the 10x36 rate map shown in panel A is plotted as a function of $n_{max}$. For $n_{max}$ values of up to 150Hz the run time is around 0.1 seconds.